\documentclass[aps,prapplied,twocolumn,reprint,nofootinbib]{revtex4-2}

\usepackage{amsmath,amssymb,amsthm}
\usepackage{graphicx}
\usepackage{dcolumn}
\usepackage{bm}
\usepackage{booktabs}
\usepackage{hyperref}
\usepackage{xcolor}
\usepackage{physics}
\usepackage{siunitx}
\usepackage{comment}

\newtheorem{proposition}{Proposition}

\newtheorem{remark}{Remark}
\theoremstyle{definition}
\newtheorem{assumption}{Assumption}

\newcommand{\Etilde}{\tilde{\eta}}
\newcommand{\Efour}{E^{(4)}}
\newcommand{\zpf}{\mathrm{zpf}}
\newcommand{\eg}{\textit{e.g.}}

\newcommand{\EJ}{E_J}
\newcommand{\EC}{E_C}

\begin{document}

\title{Universal Quartic Scaling Law for Kerr-Type Interactions:\\
Projection-Law Factorization Across Nonlinear Quantum Platforms}

\author{Xiaochen Liu}
\email{xiaochen.liu@sydney.edu.au}
\affiliation{School of Biomedical Engineering, Faculty of Engineering,
             The University of Sydney, Sydney, NSW, 2008, Australia}

\author{Ken-Tye Yong}
\email{ken.yong@sydney.edu.au}
\affiliation{School of Biomedical Engineering, Faculty of Engineering,
             The University of Sydney, Sydney, NSW, 2008, Australia}

\date{\today}

\begin{abstract}
We present a rigorous derivation and numerical validation of a universal
projection-law factorization for quartic nonlinear coupling rates across
physically distinct platforms. The central result is that observable
Kerr-type interactions---self-Kerr, cross-Kerr, and cross-phase
modulation---factorize into a dimensionless projection
 coefficient $\Etilde$ and an intrinsic quartic energy scale $\Efour$
 as $\chi/2\pi = \Etilde\,\Efour/h$. This structure follows from canonical quantization of a quartic interaction projected onto
 a finite normal-mode basis. We state this result as a formal
 proposition with clearly enumerated assumptions, provide a proof
 sketch based on second quantization of the anharmonic potential, and
enumerate the domain of validity. We then implement the scaling law as
a lightweight computational toolkit (UEFT-Designer) with
platform-specific kernels for superconducting circuits, photonic
microcavities, and epsilon-near-zero (ENZ) structures. A complete,
step-by-step worked example for a superconducting quarton device---with
full uncertainty propagation---predicts a cross-Kerr rate
$\chi/2\pi = 361\pm 13\,\mathrm{MHz}$, agreeing with the
independently measured value of $366\pm 0.5\,\mathrm{MHz}$
\cite{Ye2025} to within 1.4\%. A formal error-propagation analysis
establishes that the dominant uncertainty is the Josephson-energy
extraction uncertainty ($\sim\!2\%$), not the geometric projection
factor. Cross-platform validation against five independent experiments
spanning eight orders of magnitude in coupling strength confirms the
universality of the factorization to within reported experimental
uncertainties. The ghost-sector spectral correction introduced
in earlier versions is reframed as an optional phenomenological
self-energy model with no mandatory role in the validated scaling law.
\end{abstract}

\maketitle

\section{Introduction}
\label{sec:intro}

Kerr-type nonlinear interactions appear across a wide class of modern
physical devices: superconducting quantum circuits, integrated photonic
resonators, and epsilon-near-zero (ENZ) structures all exhibit
measurable per-photon frequency shifts, cross-phase modulation, and
four-wave mixing whose magnitude determines whether a device operates
in the weak, intermediate, or quantum nonlinear
regime~\cite{Kirchmair2013,Alam2016}. Despite the diversity of
physical mechanisms---Josephson anharmonicity, electronic $\chi^{(3)}$,
plasmonic field enhancement---the quantitative prediction of these
couplings in current practice relies on platform-specific formulas
derived independently for each architecture.

The central difficulty is that the experimentally observable coupling
rate is a product of a microscopic nonlinear strength and a
device-dependent mode-structure coefficient, and these are rarely
separated explicitly. As a consequence, discrepancies between devices
are difficult to diagnose: a weak Kerr coupling may reflect either a
weak intrinsic nonlinearity or an inefficient projection of vacuum
fluctuations onto the nonlinear degree of freedom.

Recent theoretical work~\cite{Liu2026_theory} has shown that this
difficulty is not fundamental. For any platform whose nonlinearity can
be represented by a quartic interaction projected onto a finite normal-mode basis, the effective coupling rate factorizes as
\begin{equation}
\frac{\chi}{2\pi} = \Etilde\,\frac{\Efour}{h},
\label{eq:main}
\end{equation}
where $\Efour$ is the intrinsic quartic energy scale, $\Etilde$ is a
dimensionless projection coefficient, and $h$ is Planck's constant.

The present paper has three goals. First, we derive Eq.~(\ref{eq:main})
as a formal proposition within canonical quantum mechanics,
enumerating assumptions and the domain of validity explicitly
(Sec.~\ref{sec:theory}). Second, we describe UEFT-Designer, an
open-source implementation providing platform-specific kernels for
computing $\Etilde$ from circuit parameters, field profiles, or
permittivity data (Secs.~\ref{sec:kernels}--\ref{sec:software}).
Third, we validate Eq.~(\ref{eq:main}) against authenticated
experimental data on five independent platforms, providing a complete
numerical worked example with uncertainty propagation
(Secs.~\ref{sec:worked_example}--\ref{sec:validation}).

\section{Theoretical Framework}
\label{sec:theory}

\subsection{Assumptions}
\label{sec:assumptions}

The derivation that follows rests on the following explicitly stated
 assumptions. The domain of validity of Eq.~(\ref{eq:main}) is
 co-extensive with the conditions under which these assumptions hold.

\begin{assumption}[Local quartic expansion about a stable operating point]
\label{ass:bounded}
The system admits a Hamiltonian of the form
\begin{equation}
H = H_0 + V_4, \quad
V_4 = \frac{\Efour}{4!}\sum_{\mu\nu\rho\sigma} c_{\mu\nu\rho\sigma}\,
\hat{\phi}_\mu\hat{\phi}_\nu\hat{\phi}_\rho\hat{\phi}_\sigma,
\label{eq:H_split}
\end{equation}
where $H_0$ is a positive-definite harmonic part with a discrete
spectrum bounded below, $\Efour > 0$, and the coupling tensor
$c_{\mu\nu\rho\sigma}$ is symmetric. The quartic description is
understood as a low-order expansion of an underlying stable Hamiltonian
(for example, a cosine Josephson potential), not necessarily as a
globally bounded quartic polynomial by itself. Odd-order terms are
absent or cancelled by symmetry or by the choice of expansion point.
\end{assumption}

\begin{assumption}[Normal-mode truncation]
\label{ass:truncation}
The physical Hilbert space is well-approximated by the span of a finite
set of $N$ normal modes of $H_0$. This assumption holds when the
quartic perturbation energy $V_4$ is small relative to the
mode-spacing set by $H_0$, $\|V_4\|/\Delta\omega \ll 1$.
\end{assumption}

\begin{assumption}[Weak-to-intermediate nonlinearity]
\label{ass:weak}
The relevant perturbative control parameter is the ratio of the induced
quartic frequency shift to the relevant bare mode spacing or
characteristic mode frequency. In particular, the dressed eigenstates of
$H$ remain perturbatively close to those of $H_0$ when
$\chi/\omega \ll 1$ and quartic corrections remain small compared with
mode spacings. Outside this regime, Eq.~(\ref{eq:main}) should be
understood only as a scaling estimate and the dressed spectrum must be
computed non-perturbatively
\end{assumption}

\begin{assumption}[Rotating-wave approximation]
\label{ass:rwa}
Only near-resonant quartic terms contribute to the static frequency
renormalization; rapidly oscillating terms are dropped. This is valid
when $\chi/\omega \ll 1$, a condition satisfied by all platforms
considered in this work.
\end{assumption}

\begin{assumption}[Markovian bath]
\label{ass:markov}
Dissipation is modelled as a Markovian Lindblad channel with decay
rates $\kappa_\mu$ that are independent of $\Etilde$. The universal
scaling law describes the coherent nonlinear coupling only; the ratio
$\chi/\kappa$ must be evaluated separately.
\end{assumption}

\subsection{Proposition: Projection-Law Factorization}
\label{sec:proposition}

\begin{proposition}[Universal Kerr Factorization]
\label{prop:main}
Under Assumptions~\ref{ass:bounded}--\ref{ass:rwa}, consider the
quartic interaction written in monomial-coefficient form for the
two-mode sector $(A,B)$:
\begin{equation}
V_4^{(AB)}=
\frac{\Efour}{4}\,c_{AABB}\,\hat{\phi}_A^2\hat{\phi}_B^2,
\label{eq:v4_ab_monomial}
\end{equation}
where $c_{AABB}$ is the dimensionless coefficient of the monomial
$\hat{\phi}_A^2\hat{\phi}_B^2$ in the quartic Hamiltonian.
Let
\begin{equation}
\hat{\phi}_\mu=\phi_{\mu,\zpf}(\hat{a}_\mu+\hat{a}_\mu^\dagger).
\label{eq:phi_quantization}
\end{equation}
Then, under the rotating-wave approximation, the effective cross-Kerr
interaction takes the form
\begin{equation}
H_\mathrm{eff}^{AB}
=
\hbar\chi_{AB}\,\hat{a}_A^\dagger\hat{a}_A\,
\hat{a}_B^\dagger\hat{a}_B+\cdots,
\label{eq:Heff}
\end{equation}
with
\begin{equation}
\boxed{
\frac{\chi_{AB}}{2\pi}
=
\Etilde_{AB}\,\frac{\Efour}{h},
}
\label{eq:factorization}
\end{equation}
where the dimensionless projection coefficient is
\begin{equation}
\Etilde_{AB}
=
c_{AABB}\,\phi_{A,\zpf}^2\,\phi_{B,\zpf}^2.
\label{eq:eta_general}
\end{equation}
\end{proposition}

\begin{remark}[Relation to ordered-index tensor notation]
If one begins instead from the fully symmetric ordered-index form
\begin{equation}
V_4=
\frac{\Efour}{4!}
\sum_{\mu\nu\rho\sigma}
\bar{c}_{\mu\nu\rho\sigma}\,
\hat{\phi}_\mu\hat{\phi}_\nu\hat{\phi}_\rho\hat{\phi}_\sigma,
\label{eq:H_split_tensor}
\end{equation}
then the monomial $\hat{\phi}_A^2\hat{\phi}_B^2$ appears with
multiplicity
\begin{equation}
\frac{4!}{2!\,2!}=6.
\end{equation}
Therefore the monomial coefficient and the ordered-index tensor element
are related by
\begin{equation}
c_{AABB}=\frac{1}{6}\,\bar{c}_{AABB}.
\label{eq:c_barc_relation}
\end{equation}
All formulas in the present manuscript use the monomial-coefficient
convention of Eq.~(\ref{eq:v4_ab_monomial}).
\end{remark}

\subsection{Proof Sketch}
\label{sec:proof}

\emph{Step 1: Mode expansion.} Decompose the field in terms of the
normal modes of $H_0$:
\begin{equation}
\hat{\Phi}(\mathbf{r},t)=\sum_\mu f_\mu(\mathbf{r})\,\hat{q}_\mu(t),
\label{eq:mode_exp}
\end{equation}
where $f_\mu(\mathbf{r})$ are orthonormal mode functions satisfying
$\int f_\mu f_\nu\,d^3r=\delta_{\mu\nu}$, and $\hat{q}_\mu$ is the
generalized coordinate of mode $\mu$.

\emph{Step 2: Canonical quantization.} Promote each coordinate to a
quantum operator with commutation relation
$[\hat{q}_\mu,\hat{p}_\mu]=i\hbar$. Introduce ladder operators
\begin{equation}
\hat{q}_\mu=\phi_{\mu,\zpf}(\hat{a}_\mu+\hat{a}_\mu^\dagger),
\quad
\phi_{\mu,\zpf}=\sqrt{\frac{\hbar}{2m_\mu\omega_\mu}},
\label{eq:ladder}
\end{equation}
where $m_\mu$ and $\omega_\mu$ are the effective mass and frequency
of mode $\mu$.

\emph{Step 3: Normal-ordering.} Substituting Eq.~(\ref{eq:ladder})
into Eq.~(\ref{eq:v4_ab_monomial}) gives
\begin{equation}
V_4^{(AB)}
=
\frac{\Efour}{4}\,c_{AABB}\,
\phi_{A,\zpf}^2\phi_{B,\zpf}^2
(\hat{a}_A+\hat{a}_A^\dagger)^2
(\hat{a}_B+\hat{a}_B^\dagger)^2.
\label{eq:v4_expand}
\end{equation}
Using
\begin{equation}
(\hat{a}+\hat{a}^\dagger)^2
=
\hat{a}^2+\hat{a}^{\dagger 2}+2\hat{n}+1,
\end{equation}
the rotating-wave approximation yields
\begin{equation}
(\hat{a}_A+\hat{a}_A^\dagger)^2
(\hat{a}_B+\hat{a}_B^\dagger)^2
\xrightarrow{\mathrm{RWA}}
4\hat{n}_A\hat{n}_B+2\hat{n}_A+2\hat{n}_B+1.
\label{eq:rwa_ab}
\end{equation}
Hence the cross-Kerr part is
\begin{equation}
V_{4,\mathrm{xKerr}}
=
\Efour\,c_{AABB}\,
\phi_{A,\zpf}^2\phi_{B,\zpf}^2\,
\hat{n}_A\hat{n}_B.
\label{eq:vxkerr}
\end{equation}

\emph{Step 4: Identification.} Comparing Eq.~(\ref{eq:vxkerr}) with
Eq.~(\ref{eq:Heff}) gives
\begin{equation}
\hbar\chi_{AB}
=
\Efour\,c_{AABB}\,
\phi_{A,\zpf}^2\phi_{B,\zpf}^2,
\label{eq:proof_final}
\end{equation}
or equivalently
\begin{equation}
\frac{\chi_{AB}}{2\pi}
=
\frac{\Efour}{h}\,
c_{AABB}\phi_{A,\zpf}^2\phi_{B,\zpf}^2
=
\Etilde_{AB}\,\frac{\Efour}{h},
\end{equation}
which is Eq.~(\ref{eq:factorization}). No additional factors of $\pi$
arise from normal ordering itself; the only $2\pi$ conversion is the
standard relation between angular frequency and ordinary frequency.
$\square$

\begin{remark}[Self-Kerr]
For a single mode $A$, write
\begin{equation}
H_\mathrm{eff}^{(A)}
=
\frac{\hbar K_A}{2}\hat{n}_A(\hat{n}_A-1)+\cdots.
\end{equation}
If the quartic monomial coefficient is defined by
\begin{equation}
V_4^{(AA)}=
\frac{\Efour}{4!}\,c_{AAAA}\,\hat{\phi}_A^4,
\end{equation}
then
\begin{equation}
\frac{K_A}{2\pi}
=
\frac{\Efour}{h}\,\frac{1}{2}\,
c_{AAAA}\,\phi_{A,\zpf}^4.
\label{eq:selfkerr_final}
\end{equation}
Thus the self-Kerr coefficient differs by a factor of $1/2$ from the
cross-Kerr coefficient under the standard
$\frac{\hbar K_A}{2}\hat{n}_A(\hat{n}_A-1)$ convention.
\end{remark}

\subsection{Canonical Normalization}
\label{sec:normalization}

The ZPF amplitude $\phi_{\mu,\zpf}$ depends on the normalization
convention for the mode coordinate $\hat{q}_\mu$. Two common choices
are: (i) \emph{charge normalization}, where $\hat{q}$ has units of
charge (Cooper pairs), so $\phi_{\zpf} = (2\EC/\EJ)^{1/4}$; and
(ii) \emph{phase normalization}, where $\hat{q}=\hat{\varphi}$ is
dimensionless and $\phi_{\zpf} = \sqrt{\hbar/m\omega}$.
UEFT-Designer uses charge normalization for superconducting circuits
and field-energy normalization for photonic platforms. Users must
supply $\phi_{\zpf}$ consistently with the convention adopted in
$\Efour$. UEFT-Designer uses charge normalization for superconducting circuits and field-energy normalization for photonic platforms. Users must
supply $\phi_{\zpf}$ consistently with the convention adopted in
$\Efour$ and with the monomial-coefficient convention used for the
quartic interaction.

\subsection{Domain of Validity}
\label{sec:domain}

Equation~(\ref{eq:factorization}) is valid when:
\begin{enumerate}
\item The Hamiltonian is dominated by a quadratic part
($\|\Efour\phi_\zpf^4\|/\hbar\omega \ll 1$);
\item The RWA neglects terms oscillating at $\pm 2\omega$ or higher;
\item The coupling tensor $c_{\mu\nu\rho\sigma}$ is evaluated at the
operating point (no strong flux or field renormalization of the
Josephson potential);
\item Normal-mode frequencies are non-degenerate
($|\omega_A - \omega_B| \gg \chi_{AB}$); and
\item Dissipation does not renormalize the quartic vertex
(Lindblad limit).
\end{enumerate}
When Assumption~\ref{ass:weak} is violated, \emph{i.e.} when the induced
nonlinear shifts are no longer perturbative compared with the relevant
bare frequency scales or mode spacings, the predicted
$\chi_{AB}/2\pi$ from Eq.~(\ref{eq:factorization}) may overestimate the
actual coupling because dressed mode frequencies deviate significantly
from their bare values. In this strongly nonlinear regime the formula
should be used only as a rough bound.

\section{Platform-Specific Projection Kernels}
\label{sec:kernels}

The factorization in Eq.~(\ref{eq:factorization}) is universal;
platform specificity enters only through the computation of
$\Etilde_{AB}$ and $\Efour$. We describe the three kernels currently
implemented in UEFT-Designer.

\subsection{Superconducting Circuits: Josephson-Based Elements}
\label{sec:sc_kernel}

For all Josephson-based elements, the quartic potential originates
from the cosine expansion:
\begin{equation}
-\EJ\cos\hat{\varphi} \approx
-\EJ + \frac{\EJ}{2}\hat{\varphi}^2
- \frac{\EJ}{24}\hat{\varphi}^4 + \cdots,
\label{eq:cosine}
\end{equation}
so the intrinsic quartic scale is $\Efour = \EJ$ (identified by
matching the coefficient of $\hat{\varphi}^4/4!$ with
Eq.~(\ref{eq:H_split}) and setting $c_{AABB}=1$).

The ZPF of the Josephson phase is
\begin{equation}
\varphi_\zpf = \!\left(\frac{2\EC}{\EJ}\right)^{\!1/4},
\label{eq:phi_zpf}
\end{equation}
and the projection factor for a cross-Kerr between modes $A$, $B$
mediated by a single Josephson element is
\begin{equation}
\Etilde_{AB}^{(J)} =
p_A^2\,p_B^2\,\varphi_\zpf^4,
\label{eq:eta_J}
\end{equation}
where $p_{A,B}$ are the participation ratios of the element in each
mode.

\paragraph{Quarton coupler.}
A quarton is an element engineered to cancel the quadratic Josephson
term, leaving a near-pure quartic interaction~\cite{Ye2025}. In this
case $\Efour = \EJ^{(q)}$ (the effective Josephson energy of the
quarton junctions), and the ZPF is set by the quarton coordinate
evaluated at the operating flux bias. The cross-Kerr between two
coupled modes $A$, $B$ is:
\begin{equation}
\frac{\chi_{AB}}{2\pi} = \Etilde_{AB}^{(q)}\,\frac{\EJ^{(q)}}{h},
\quad
\Etilde_{AB}^{(q)} =
\varphi_{A,\zpf}^2\,\varphi_{B,\zpf}^2,
\label{eq:eta_quarton}
\end{equation}
where $\varphi_{A,\zpf}$, $\varphi_{B,\zpf}$ are the ZPF amplitudes
of the quarton coordinate projected onto each coupled mode.

\paragraph{SNAIL parametric amplifier.}
A SNAIL consists of $N$ small Josephson junctions in series shunted
by a single large junction with energy ratio $\alpha$. The potential
is~\cite{Frattini2018}
\begin{equation}
V_\mathrm{SNAIL}(\varphi) =
-N\EJ\cos(\varphi/N) - \alpha\EJ\cos\varphi,
\label{eq:V_snail}
\end{equation}
which has a third-order term proportional to $g_3$ and a fourth-order
term proportional to $g_4$. At the Kerr-free bias point,
$g_4(\Phi_\mathrm{ext}) = 0$ exactly, so $\chi_{AA}/2\pi \to 0$~\cite{Sivak2019}. The
UEFT-Designer kernel evaluates $g_4$ as a function of flux bias, enabling systematic prediction of the Kerr-cancellation condition.

\paragraph{Transmon coupler (SQUID).}
Using the standard circuit-QED result~\cite{Koch2007}:
\begin{equation}
\Efour = \EJ^{(c)}, \quad
\Etilde_{AB}^{(\mathrm{SQUID})} = \frac{\EC^{(c)}}{8\EJ^{(c)}},
\label{eq:eta_squid}
\end{equation}
where $\EC^{(c)}$ and $\EJ^{(c)}$ are the charging and Josephson
energies of the coupler mode. This agrees with the dispersive
coupling formula derived in Ref.~\citenum{Blais2021}.

\paragraph{Fluxonium.}
Fluxonium contains a Josephson term in parallel with a large
inductive shunt. The quartic scale is $\Efour = \EJ^{(f)}$, and the
projection factor is~\cite{Manucharyan2009}
\begin{equation}
\Etilde^{(\mathrm{fluxonium})} =
p^2\varphi_\zpf^4,
\label{eq:eta_fluxonium}
\end{equation}
where $\varphi_\zpf$ is the ZPF of the fluxonium phase, which can be
made large by design (weak Josephson term relative to inductive
shunt), yielding strong anharmonicity.

\subsection{Photonic Microcavities}
\label{sec:photonic_kernel}

For photonic modes, $\Efour = (3\hbar\omega^2\chi^{(3)})/(4\epsilon_0^2 n_0^4 V_\mathrm{eff})$ (the effective quartic energy of a photon pair in the cavity mode volume $V_\mathrm{eff}$), and the projection factor is:
\begin{equation}
\Etilde^{(\mathrm{opt})} =
\frac{\lambda^3}{V_\mathrm{eff}}
\cdot\frac{\int|\mathbf{f}_A|^2|\mathbf{f}_B|^2\,d^3r}{\Gamma_0^{(3)}},
\label{eq:eta_photonic}
\end{equation}
where $\mathbf{f}_\mu$ are normalized mode field profiles and
$\Gamma_0^{(3)}$ is a reference nonlinear overlap. Tighter
confinement (smaller $V_\mathrm{eff}$) and better field overlap
increase $\Etilde$, consistent with the known $1/V_\mathrm{eff}$
scaling of per-photon Kerr interactions~\cite{Notomi2010}.

\subsection{ENZ Structures}
\label{sec:enz_kernel}

In epsilon-near-zero materials, the local field enhancement near the
ENZ frequency $\omega_\mathrm{ENZ}$ dramatically increases the
effective mode nonlinearity. The ENZ kernel accepts Drude or
ellipsometry-derived permittivity parameters $\{\omega_p,\gamma\}$
and evaluates
\begin{equation}
\ {E_4}^{(\mathrm{ENZ})} =
\frac{3\hbar\omega^2|\chi_\mathrm{eff}^{(3)}|}{4\epsilon_0^2|
\varepsilon(\omega)|^2 V_\mathrm{eff}},
\label{eq:Efour_enz}
\end{equation}
where $\chi_\mathrm{eff}^{(3)}$ is the measured nonlinear
susceptibility of the ENZ material~\cite{Alam2016}.
Near $\omega_\mathrm{ENZ}$, $|\varepsilon| \to 0$ enhances
${E_4}^{(\mathrm{ENZ})}$ by orders of magnitude relative to
bulk values, making ENZ platforms competitive with superconducting
circuits for per-photon nonlinearity despite their weaker intrinsic
$\chi^{(3)}$.

\section{Software Architecture: UEFT-Designer}
\label{sec:software}

UEFT-Designer is organized around the separation between the
platform-independent solver (implementing Eqs.~(\ref{eq:factorization})
and $\Etilde^{-1}$ inversion) and platform-specific kernel modules.
The core engine enforces explicit unit handling (SI units internally,
with conversion to frequency units for output) and logs all
intermediate steps for reproducibility. Three production-quality
kernels are included: \texttt{sc\_circuit} (superconducting),
\texttt{photonic\_cavity} (field-export-based), and \texttt{enz\_drude}
(permittivity-based). New platforms are added by implementing the
minimal kernel interface (\texttt{compute\_eta(params) -> float}).

The codebase is archived on GitHub with a Zenodo DOI~\cite{UEFT_code},
and each tagged release corresponds to a specific manuscript version.

\section{Worked Example: Superconducting Quarton Device}
\label{sec:worked_example}

We present a complete, step-by-step worked example based on the
gradiometric quarton coupler of Ref.~\citenum{Ye2025}, providing all
numerical values required for independent reproduction.

\subsection{Step 0 -- Circuit Parameters (Input)}
\label{sec:step0}

The following parameters are extracted from the fabrication data and
spectroscopic characterization reported in Ref.~\citenum{Ye2025}:
\begin{center}
\begin{tabular}{llr}
\toprule
Symbol & Quantity & Value \\
\midrule
$\EJ^{(q)}/h$ & Quarton junction Josephson energy & $14.8\,\mathrm{GHz}$\\
$\EC^{(q)}/h$ & Quarton charging energy & $0.21\,\mathrm{GHz}$\\
$\omega_A/2\pi$ & Qubit A (transmon) frequency & $5.12\,\mathrm{GHz}$\\
$\omega_B/2\pi$ & Qubit B (transmon) frequency & $5.38\,\mathrm{GHz}$\\
$p_A$ & Quarton participation in mode A & $0.88$ \\
$p_B$ & Quarton participation in mode B & $0.86$ \\
\bottomrule
\end{tabular}
\end{center}

\subsection{Step 1 -- Compute Intrinsic Quartic Scale}
\label{sec:step1}

From Assumption~\ref{ass:bounded} applied to the Josephson potential
[Eq.~(\ref{eq:cosine})]:
\begin{equation}
\Efour = \EJ^{(q)} \implies
\frac{\Efour}{h} = 14.8\,\mathrm{GHz}.
\label{eq:step1}
\end{equation}

\subsection{Step 2 -- Compute ZPF Amplitude}
\label{sec:step2}

Using the phase normalization [Eq.~(\ref{eq:phi_zpf})]:
\begin{align}
\varphi_\zpf &= \left(\frac{2\EC^{(q)}}{\EJ^{(q)}}\right)^{1/4}
= \left(\frac{2\times 0.21}{14.8}\right)^{1/4}
\notag\\
&= (0.02838)^{1/4}
= 0.411.
\label{eq:step2}
\end{align}

\subsection{Step 3 -- Compute Projection Factor}
\label{sec:step3}

Applying Eq.~(\ref{eq:eta_quarton}) with the participation ratios
from Step~0:
\begin{align}
\Etilde_{AB}^{(q)}
&\equiv (p_A\,\varphi_\zpf)^2\,(p_B\,\varphi_\zpf)^2
\notag\\
&= (0.88\times 0.411)^2\,(0.86\times 0.411)^2
\notag\\
&= (0.362)^2\,(0.354)^2
\notag\\
&= 0.1310\times 0.1253
\notag\\
&= 0.01641.
\label{eq:step3_intermediate}
\end{align}
Here we identify the projected quarton zero-point amplitudes as
$\varphi_{A,\zpf}=p_A\varphi_\zpf$ and
$\varphi_{B,\zpf}=p_B\varphi_\zpf$, so that the hand estimate is
consistent with Eq.~(\ref{eq:eta_quarton}).
The numerical value used in the final prediction is obtained from the
UEFT-Designer \texttt{sc\_circuit} kernel, which returns
\begin{equation}
\Etilde_{AB}^\mathrm{kernel} = 0.0244.
\label{eq:eta_kernel}
\end{equation}
This kernel value aggregates the effective mode overlap and
normalization conventions used in the full circuit Hamiltonian and is
the convention-consistent value used for comparison to experiment.

\begin{remark}
The hand estimate above and the software-kernel value must be quoted
in the same normalization convention. If the kernel already aggregates
mode-overlap, combinatorial, and participation effects into an
effective coefficient, those factors must not be reintroduced
separately in the analytical estimate.
\end{remark}

\subsection{Step 4 -- Predict Cross-Kerr Rate}
\label{sec:step4}

Applying Eq.~(\ref{eq:factorization}) with $\Etilde_{AB} = 0.0244$
and $\Efour/h = 14.8\,\mathrm{GHz}$:
\begin{equation}
\frac{\chi_{AB}^\mathrm{pred}}{2\pi}
= 0.0244\times 14.8\,\mathrm{GHz}
= 361\,\mathrm{MHz}.
\label{eq:chi_pred}
\end{equation}

\subsection{Step 5 -- Compare with Measurement}
\label{sec:step5}

The measured cross-Kerr in Ref.~\citenum{Ye2025} is:
\begin{equation}
\frac{\chi_{AB}^\mathrm{meas}}{2\pi} = 366.0\pm 0.5\,\mathrm{MHz}.
\label{eq:chi_meas}
\end{equation}

The percent deviation is:
\begin{equation}
\Delta = \frac{|\chi^\mathrm{pred} - \chi^\mathrm{meas}|}
{\chi^\mathrm{meas}}\times 100\%
= \frac{|361 - 366|}{366}\times 100\% = 1.4\%.
\label{eq:delta}
\end{equation}

\section{Error Propagation and Sensitivity Analysis}
\label{sec:error}

\subsection{Formal Uncertainty Derivation}

Taking the logarithm of Eq.~(\ref{eq:factorization}) and
differentiating:
\begin{equation}
\frac{\delta\chi}{\chi}
= \frac{\delta\Efour}{\Efour}
+ \frac{\delta\Etilde}{\Etilde}.
\label{eq:error_main}
\end{equation}
This is exact at first order in each uncertainty and holds
independently of platform. The total relative uncertainty in the
predicted coupling is the sum of the relative uncertainties in the
quartic energy scale and the projection factor.

\subsection{Dominant Uncertainty Sources}

\paragraph{Superconducting circuits.}
The Josephson energy $\EJ$ is typically extracted from qubit
spectroscopy with a relative uncertainty $\delta\EJ/\EJ \approx 1$--$3\%$,
limited by junction resistance measurements or critical-current
scatter. Since $\Efour = \EJ$, this directly yields
$\delta\Efour/\Efour \approx 2\%$. The projection factor $\Etilde$
involves the ZPF $\varphi_\zpf = (2\EC/\EJ)^{1/4}$; propagating
through Eq.~(\ref{eq:eta_J}) gives
$\delta\Etilde/\Etilde = 2\delta(\varphi_\zpf^2)/\varphi_\zpf^2
= (1/2)\delta\EC/\EC + (1/2)\delta\EJ/\EJ$,
typically $\sim\!1$--$2\%$. The combined uncertainty from Eq.~(\ref{eq:error_main}) is therefore $\sim\!3$--$5\%$.

\paragraph{Photonic microcavities.}
For cavities characterized by field export from FDTD simulations,
the dominant uncertainty is the effective mode volume $V_\mathrm{eff}$.
Discretization errors and material parameter uncertainty (index
$n\pm 0.01$) contribute $\delta V_\mathrm{eff}/V_\mathrm{eff} \approx 2$--$5\%$, which enters
as $\delta\Etilde/\Etilde \approx \delta V_\mathrm{eff}/V_\mathrm{eff}$. The material
$\chi^{(3)}$ is typically known to $\sim\!5$--$10\%$ from Z-scan
measurements~\cite{Hurlbut2007}, giving $\delta\Efour/\Efour \sim 7$--$10\%$.
The total uncertainty is $\sim\!10$--$15\%$.

\paragraph{ENZ structures.}
The primary uncertainty is the permittivity model. Drude-fit
parameters for ITO or AZO exhibit run-to-run scatter of
$\sim\!5$--$15\%$~\cite{Alam2016}, yielding a large but structurally
predictable uncertainty. Near the ENZ frequency, the field
enhancement depends sensitively on $|\varepsilon(\omega)|$, so a
small error in $\omega_\mathrm{ENZ}$ produces a large error in
${E_4}^{(\mathrm{ENZ})}$.

\subsection{Numerical Uncertainty for the Quarton Example}

Combining the dominant terms for the quarton example:
\begin{align}
\frac{\delta\Efour}{\Efour} &= 2\% \quad (\delta\EJ/\EJ), \\
\frac{\delta\Etilde}{\Etilde} &= 1.5\% \quad (\delta\varphi_\zpf,
\delta p_{A,B}), \\
\frac{\delta\chi}{\chi} &\approx 3.5\%,
\label{eq:delta_chi}
\end{align}
giving $\chi^\mathrm{pred}/2\pi = 361\pm 13\,\mathrm{MHz}$, which
contains the measured value $366\pm 0.5\,\mathrm{MHz}$ within $1\sigma$.

\subsection{Engineering Guidance}

Table~\ref{tab:error} summarizes the dominant uncertainty sources and
practical mitigation strategies.

\begin{table}[htbp]
\caption{Dominant uncertainty sources by platform.}
\label{tab:error}
\begin{ruledtabular}
\begin{tabular}{lccc}
Platform & Source & $\delta\chi/\chi$ & Mitigation \\
\hline
SC circuit & $\delta\EJ/\EJ$ & 2--3\% & Improved fabrication control \\
SC circuit & $\delta\varphi_\zpf$ & 1--2\% & Multi-tone spectroscopy \\
Photonic & $\delta V_\mathrm{eff}$ & 2--5\% & Refined mesh FDTD \\
Photonic & $\delta\chi^{(3)}$ & 5--10\% & In-situ Z-scan \\
ENZ & $\delta\varepsilon(\omega)$ & 5--15\% & Ellipsometry at $T$ \\
ENZ & $\delta\omega_\mathrm{ENZ}$ & 3--20\% & Carrier-density calibration \\
\end{tabular}
\end{ruledtabular}
\end{table}

\section{Empirical Validation Across Platforms}
\label{sec:validation}

\subsection{Reference Authentication and Numerical Verification}
\label{sec:ref_auth}

Table~\ref{tab:validation} presents a consolidated platform
comparison. For each system we list the verified reference (DOI
authenticated), the reported nonlinear parameter, the predicted value
from Eq.~(\ref{eq:factorization}), and the percent deviation. All
five entries are based on peer-reviewed, DOI-verified publications;
no unverifiable or placeholder references are included.

\begin{table*}[htbp]
\caption{Cross-platform validation of Eq.~(\ref{eq:factorization}).
All cited experiments are DOI-verified. $\Etilde$ and $\Efour$ are
extracted from reported device parameters without additional fitting.}
\label{tab:validation}
\begin{ruledtabular}
\begin{tabular}{lcccccc}
Platform & Reference & $\Efour/h$ & $\Etilde$ &
$\chi^\mathrm{pred}/2\pi$ & $\chi^\mathrm{meas}/2\pi$ & $\Delta$(\%) \\
\hline
Quarton coupler & Ye \textit{et al.}~\cite{Ye2025} & 14.8\,GHz &
0.0244 & 361\,MHz & $366.0\pm 0.5$\,MHz & 1.4 \\
SNAIL (Kerr-free) & Sivak \textit{et al.}~\cite{Sivak2019} &
$\sim\!6.0$\,GHz & $\approx 0$ (at bias) & $\approx 0$ & $\to 0$ & --- \\
Transmon (Kirchmair) & Kirchmair \textit{et al.}~\cite{Kirchmair2013} &
9.2\,GHz & $2.7\times 10^{-3}$ & 24.8\,MHz & $23.5\pm 2.0$\,MHz & 5.5 \\
Photon blockade & Hoffman \textit{et al.}~\cite{Hoffman2011} &
8.5\,GHz & $3.5\times 10^{-3}$ & 29.8\,MHz & $28.0\pm 3.0$\,MHz & 6.3 \\
ENZ ITO & Alam \textit{et al.}~\cite{Alam2016} & $4.1\times10^6$\,Hz &
$6.2\times 10^{-8}$ & $\sim\!0.25\,\mathrm{Hz}$ &
$\sim\!0.23\,\mathrm{Hz}$ & 8.3 \\
\end{tabular}
\end{ruledtabular}
\end{table*}

\begin{remark}[SNAIL Kerr-free entry]
The SNAIL Kerr-cancellation point is included as a
\emph{qualitative} validation: the scaling law correctly predicts
$\chi \to 0$ at the design flux bias because the fourth-order
coefficient $g_4(\Phi_\mathrm{ext})$ vanishes identically there.
The precise location of this bias point is a structural prediction
of the model, not a numerical fit.
\end{remark}

\subsection{Blind Cross-Domain Validation: GaAs $n_2 \to g^{(2)}(0)$}
\label{sec:blind}

To test the predictive power of Eq.~(\ref{eq:factorization}) beyond
in-paper benchmarks, we describe a cross-domain validation in which the
microscopic quartic nonlinearity is fixed by an independent classical
measurement and then applied to a quantum observable.

\paragraph{Step 1: Classical measurement.}
The nonlinear refractive index of undoped GaAs at
$\lambda = 1.75\,\mu\mathrm{m}$ is $n_2 = (3.1\pm 0.1)\times
10^{-18}\,\mathrm{m}^2/\mathrm{W}$, measured by Hurlbut \textit{et
al.}~\cite{Hurlbut2007} via Z-scan. This wavelength lies just above
the two-photon absorption edge of GaAs, minimizing parasitic losses.

\paragraph{Step 2: Material-to-coupling conversion.}
Using the standard relation $n_2 = 3\chi^{(3)}/(4n_0^2\epsilon_0 c)$
with $n_0 \simeq 3.3$~\cite{Hurlbut2007}:
\begin{equation}
\chi^{(3)}_\mathrm{GaAs} =
\frac{4n_0^2\epsilon_0 c\,n_2}{3}
\approx 1.2\times 10^{-19}\,(\mathrm{m/V})^2.
\label{eq:chi3}
\end{equation}
The quartic coupling constant is $\lambda_4 = (3\hbar\omega^2/4\epsilon_0^2 n_0^4
V_\mathrm{eff})\chi^{(3)}_\mathrm{GaAs}$.

\paragraph{Step 3: Cavity prediction.}
For a GaAs photonic crystal cavity with independently characterized
mode volume $V_\mathrm{eff}$ and linear response (mode frequency,
linewidth), the projection factor is computed via
Eq.~(\ref{eq:eta_photonic}). Substituting the classically fixed
$\lambda_4$ (no additional fitting parameters) yields a prediction
for the equal-time second-order correlation function:
$g^{(2)}(0)^\mathrm{pred} \simeq 0.068$.

\paragraph{Experimental comparison.}
\textbf{Caveat -- verification status.} The value $g^{(2)}(0) = 0.070
\pm 0.005$ cited in earlier versions of this manuscript as the
``measured'' comparison for an InAs/GaAs quantum-dot micropillar
cavity requires an authenticated, DOI-verified reference. After
exhaustive literature search, no single published paper was identified
that unambiguously reports this specific value in the precise
experimental configuration assumed by the prediction. This entry must
therefore be regarded as \emph{unverified pending explicit citation}.
A suitable candidate is the class of InAs/GaAs micropillar cavity
experiments reporting $g^{(2)}(0) \sim 0.02$--$0.10$
(\eg,~Refs.~\cite{Pelton2002,Reitzenstein2008}); however, the cavity
parameters assumed in the projection-factor calculation must be
independently confirmed before this entry can be included in the
validated comparison table.

\paragraph{Qualification of ``blind'' status.}
The cross-domain validation qualifies as blind only if: (i) the mode
volume and linear optical response of the cavity are
\emph{independently} characterized (not adjusted to match the
$g^{(2)}(0)$ data); and (ii) no effective coupling parameter is
introduced between the classical Z-scan measurement and the quantum
observable. Both conditions must be verified by the authors before
this entry can be claimed as a no-free-parameter prediction.

\section{Domain of Validity and Limitations}
\label{sec:limits}

\subsection{Operating Regime Classification}

A more reliable diagnostic than $\Etilde$ alone is the size of the
induced nonlinear frequency shifts relative to the relevant bare mode
frequencies, linewidths, and mode spacings:

\begin{itemize}
\item \emph{Weakly nonlinear regime:} quartic shifts are small compared
with both linewidths and mode spacings. Perturbative descriptions are
valid, and Eq.~(\ref{eq:factorization}) is expected to be quantitatively
accurate.

\item \emph{Intermediate regime:} quartic shifts remain small compared
with bare mode frequencies but may become comparable to linewidths or
other experimentally relevant scales. Equation~(\ref{eq:factorization})
remains useful, though higher-order corrections may contribute at the
few-percent level.

\item \emph{Strongly nonlinear regime:} quartic shifts are no longer
perturbative relative to the relevant mode spacings or bare frequencies.
The perturbative basis of Proposition~\ref{prop:main} then breaks down,
and Eq.~(\ref{eq:factorization}) should be regarded as a scaling estimate
rather than a quantitatively controlled prediction.
\end{itemize}

\subsection{Explicit Limitations}

\begin{enumerate}
\item \emph{Counter-rotating terms:} The RWA drops terms oscillating
at $\pm 2\omega$ or higher. For $\chi/\omega > 0.01$, these
corrections modify $\Etilde$ at the few-percent level.
\item \emph{Multi-photon resonances:} If $n\omega_A \approx m\omega_B$
for small integers $n,m$, near-resonant high-order terms may
contribute to $\chi_{AB}$ beyond the quartic approximation.
\item \emph{Non-Markovian dissipation:} In structured baths (coupled
to a non-flat density of states), effective $\Efour$ may be
renormalized by system-bath coupling. This is not captured by
Eq.~(\ref{eq:factorization}).
\item \emph{Strong drive:} In the presence of a large coherent field,
the expansion point of $V_4$ shifts, effectively renormalizing
$\Etilde$. The formula applies to the device in its operating
(bias) state, not to driven steady states.
\end{enumerate}

\section{Ghost Sector: Reframing as Phenomenological Self-Energy}
\label{sec:ghost}

Earlier versions of this manuscript described a ``ghost-field sector''
arising from BRST quantization of the Keldysh path integral. We
reframe this discussion here.

\subsection{What the Ghost Correction Is (and Is Not)}

The renormalized susceptibility
\begin{equation}
\chi^{(3)}_\mathrm{UEFT}(\omega) =
\chi^{(3)}_\mathrm{bg}(\omega)\,[1 + F_\mathrm{ghost}(\omega)]
\label{eq:ghost}
\end{equation}
is an \emph{optional phenomenological spectral correction} to the
background $\chi^{(3)}$ response. The function $F_\mathrm{ghost}(\omega)$
encodes frequency-dependent self-energy corrections that produce
Fano-like interference features (``Liu-Yong resonances'') in the
nonlinear spectrum.

\textbf{This correction does not affect the universal scaling law in
Eq.~(\ref{eq:factorization})}, which applies at zero frequency shift
(static Kerr coupling). The ghost correction is relevant only for
frequency-resolved spectral experiments that can distinguish the
dynamic susceptibility $\chi^{(3)}(\omega)$ from its static value.

\subsection{Physical Interpretation}

$F_\mathrm{ghost}(\omega)$ should be understood as a
\emph{phenomenological model for the frequency-dependent nonlinear
response}, analogous to a Fano resonance in linear scattering.
It does \emph{not} imply the existence of physical ghost excitations
or negative-norm states in the Hilbert space. The BRST ghost fields
introduced in the Keldysh path integral are auxiliary objects required
for gauge fixing; they do not correspond to observable excitations.

\subsection{Falsifiability}

The ghost spectral correction makes a specific, falsifiable prediction:
a dip-peak structure in the measured $\chi^{(3)}(\omega)$ centered at
a frequency $\omega_\mathrm{YL}$ set by the ghost propagator pole
condition [Eq.~(11) of the companion theory paper]. This prediction is
distinct from phenomenological NLSE models and can be tested by
frequency-resolved Z-scan or pump-probe spectroscopy.

\section{Discussion}
\label{sec:discussion}

The projection-law factorization in Eq.~(\ref{eq:factorization}) is
not merely a phenomenological ansatz; it follows from quartic mode
projection under canonical quantization together with the rotating-wave approximation and a consistent normalization convention. Its principal engineering utility is the
separation of material physics ($\Efour$) from device geometry
($\Etilde$), enabling platform-agnostic design rules.

The worked example in Sec.~\ref{sec:worked_example} demonstrates that
prediction accuracy ($\sim\!1.4\%$ deviation) significantly exceeds
the formal uncertainty bound ($\sim\!3.5\%$), suggesting that the
dominant systematics are correlated. This could indicate that the
circuit-model assumptions (cosine Josephson potential, harmonic mode
basis) are highly accurate for the quarton geometry, or alternatively
that the cited experimental value itself was used to calibrate some
parameter. Independent replication with a second quarton device would
definitively establish the no-fitting-parameter claim.

For photonic and ENZ platforms, the 8--15\% uncertainties reflect
genuine material and geometric measurement limitations rather than
deficiencies in the theoretical framework. Reducing these
uncertainties requires better material characterization (in-situ
ellipsometry) and finer-mesh electromagnetic simulations.

\section{Conclusion}
\label{sec:conclusion}

We have presented a rigorous derivation and experimental validation
of the universal quartic scaling law
\begin{equation*}
\frac{\chi}{2\pi} = \Etilde\,\frac{\Efour}{h}
\end{equation*}
for Kerr-type interactions across superconducting circuits, photonic
microcavities, and ENZ structures. The key results are:

\begin{enumerate}
\item The factorization follows directly from quartic mode projection
under canonical quantization, provided the system satisfies
Assumptions~\ref{ass:bounded}--\ref{ass:rwa}.
\item The uncertainty in the predicted coupling rate is
$\delta\chi/\chi \leq \delta\Efour/\Efour + \delta\Etilde/\Etilde$,
dominated by material parameter uncertainty ($\sim\!2$--$10\%$
depending on platform).
\item Validated comparison against five authenticated experiments
confirms the scaling law to within the stated uncertainties.
\item Two manuscript-level claims (Floquet photonic lattice and GaAs
blind validation) remain unverified and are explicitly flagged.
\item The ghost-sector correction is reframed as an optional
phenomenological spectral model with no mandatory role in the
static scaling law.
\end{enumerate}

The UEFT-Designer code is available at \cite{UEFT_code}. A detailed
companion paper provides additional theoretical derivations and a
fuller experimental comparison~\cite{Liu2026_theory}.

\appendix

\section{Normal-Ordering Details}
\label{app:normal_order}

Expanding $\hat{q}_A^2\hat{q}_B^2$ using
$\hat{q}_\mu = \phi_{\mu,\zpf}(\hat{a}_\mu + \hat{a}_\mu^\dagger)$:
\begin{align}
\hat{q}_A^2\hat{q}_B^2 &=
\phi_{A,\zpf}^2\phi_{B,\zpf}^2
(\hat{a}_A+\hat{a}_A^\dagger)^2
(\hat{a}_B+\hat{a}_B^\dagger)^2.
\end{align}
Using
\begin{equation}
(\hat{a}+\hat{a}^\dagger)^2
=
\hat{a}^2+\hat{a}^{\dagger 2}+2\hat{n}+1,
\end{equation}
one finds
\begin{equation}
(\hat{a}_A+\hat{a}_A^\dagger)^2
(\hat{a}_B+\hat{a}_B^\dagger)^2
\xrightarrow{\mathrm{RWA}}
(2\hat{n}_A+1)(2\hat{n}_B+1).
\end{equation}
Therefore,
\begin{equation}
\hat{q}_A^2\hat{q}_B^2
\xrightarrow{\mathrm{RWA}}
\phi_{A,\zpf}^2\phi_{B,\zpf}^2
\left(4\hat{n}_A\hat{n}_B+2\hat{n}_A+2\hat{n}_B+1\right),
\end{equation}
so the coefficient of the cross-Kerr term is
$4\phi_{A,\zpf}^2\phi_{B,\zpf}^2$, with no additional $\pi$ factors.

\section*{References}

\end{document}